\documentstyle[11pt]{article}

    \advance\textheight by \topskip

\begin{document}

\begin{center}
{\huge \bf The method of direct expansions of Feynman integrals} \\[10mm] 
  S.A. Larin \\ [3mm]
 Institut f\"ur Theoretische Teilchenphysik, Universit\"at Karlsruhe,\\
 D-76128 Karlsruhe, Germany\\
 and\\
 Institute for Nuclear Research of the
 Russian Academy of Sciences,   \\
 60th October Anniversary Prospect 7a,
 Moscow 117312, Russia
\end{center}

\vspace{30mm}

\begin{abstract}
The universal method of expansion of integrals is
suggested. 
It allows in particular to derive  
the threshold expansion of Feynman integrals.
\end{abstract}

\newpage

During the last two decades the efficient techniques were 
developed \cite{asex1}-\cite{la}
for expansions of Feynman integrals in external parameters.
These techniques allow to perform
practical analytic calculations
in quantum field theory when exact integrations are not possible.

Recently the prescription was suggested \cite{bs} for
expansions of Feynman integrals near thresholds
but its consistent derivation  is missing,
see also \cite{s}.

In the present paper we suggest
the universal method of expansion of integrals 
 -- 'the method of direct expansions'. In particular it
allows to derive consistently the threshold expansion.

To regularize integrals we will use dimensional
regularization \cite{dr} for convenience (but the method
is regularization independent). 
The dimension of the momentum space is 
$D=4-2\epsilon$ where $\epsilon$ is the parameter 
defining the deviation of the dimension from its physical 
value 4.

To demonstrate the essence
of the method of direct expansions we begin with the following
simple integral depending on some external parameter $t$
\begin{equation}
\label{examplea}
I_1 \equiv \int_{0}^{1}\frac{dx}{(x+t)^{1+\epsilon}}
=\frac{1}{\epsilon}\frac{1}{t^{\epsilon}}
-\frac{1}{\epsilon}\frac{1}{(1+t)^{\epsilon}},
\end{equation}
where $\epsilon$ is the parameter of dimensional regularization.

We want to expand this integral in small $t$ before the integration.
Let us directly expand the integrand  in small $t$
\begin{equation}
\label{exampleb}
\int_{0}^{1} dx \left[ \frac{1}{x^{1+\epsilon}}-
(1+\epsilon)\frac{1}{x^{2+\epsilon}}t+
(1+\epsilon)(2+\epsilon)\frac{1}{x^{3+\epsilon}}\frac{t^2}{2}
+O(t^3)\right]=
\end{equation}
\[
-\frac{1}{\epsilon} +t -(1+\epsilon)\frac{t^2}{2}+
 O(t^3).
\]
In the absence of dimensional regularization, i.e. for $\epsilon=0$,
this expansion would produce non-integrable singularities at $x=0$.
But dimensional regularization makes these singularities integrable and
this direct expansion correctly reproduces
the small $t$ expansion of the exact result in the right hand side of 
eq.(\ref{examplea}). Only the term
$\frac{1}{\epsilon}\frac{1}{t^{\epsilon}}$ is invisible
for the small $t$ expansion (i.e. its expansion gives zero in any order) 
due to the basic property of dimensional regularization
\begin{equation}
\label{drprop}
 \frac{1}{t^{\epsilon+n}}=0~~~ for~~~ t=0~~~ or~~~ t=\infty,~~~n \neq-\epsilon,
\end{equation}
where $n$ can be an arbitrary number.

One can say that the term
$\frac{1}{\epsilon}\frac{1}{t^{\epsilon}}$ is a non-perturbative term
for the small $t$ expansion. Thus the expansion (\ref{exampleb})
is correct but not complete, since it has the non-perturbative
term which can be not small
for small $t$, e.g. its expansion in $\epsilon$ gives $ln(t)$.

Let us now expand the integral $I_1$ in large $t$.
We again directly expand the integrand. This means
that one factorizes the factor
$\frac{1}{t^{1+\epsilon}}$ and then expands the integrand
in a Taylor series in $1/t$
or, in other words, one expands the integrand in small $x$
\begin{equation}
\label{examplec}
\int_{0}^{1} dx \left[\frac{1}{t^{1+\epsilon}}-
(1+\epsilon)\frac{1}{t^{2+\epsilon}}x+
(1+\epsilon)(2+\epsilon)\frac{1}{t^{3+\epsilon}}\frac{x^2}{2}+O(x^3)
\right]=
\end{equation}
\[
\frac{1}{t^{1+\epsilon}} -(1+\epsilon)\frac{1}{2t^{2+\epsilon}}+
(1+\epsilon)(2+\epsilon)\frac{1}{6t^{3+\epsilon}}+
O\left(\frac{1}{t^{4+\epsilon}}\right).
\]
Again this expansion correctly reproduces
the expansion in large $t$ of the exact result in the r.h.s. of 
(\ref{examplea}). The expansion in large $t$ for the exact result means
that one first factorizes the non-perturbative factor
$\frac{1}{t^{\epsilon}}$ and then expand the rest in the Taylor series
in $1/t$.
The expansion (\ref{examplec}) is complete, i.e. there are no
non-perturbative terms for it.
This is because the terms of this expansion are integrable
even for $\epsilon=0$, i.e. in the absence of dimensional regularization.

Let us generalize the above considerations.
This simple example reflects the principal property of expansions
of integrals: an expansion of a given integrand produces a
correct result if all terms of this expansion are integrable.
(For the integral $I_1$
 it is dimensional regularization which makes all
terms of expansions integrable. For some integrals it can  happen
with the help of other regularizations, e.g. analytic regularization.)
A correct result means that it correctly reproduces a corresponding
part of an exact result for a given integral.
But the essential point is that a correct result does not always
mean a complete result.
There can be non-perturbative terms in an exact result
which are invisible for some expansions, i.e. expansions of these
terms give zero in any order.
(In the above example it was the term $1/t^{\epsilon}$ which was invisible
in any order of the expansion in small $t$.) 
To get a complete expansion of a given integral
one should construct a complete set of direct
expansions of an integrand to reproduce all necessary terms.

To see how the method works practically let us apply it to
a Feynman integral. 
We will consider the following two loop propagator Feynman
integral with an external momentum $q$ and a mass $M$
\begin{equation}
\label{largeM1}
I_2 \equiv \int\frac{d^Dk~d^Dl}
{(k-q)^2k^2[(k-l)^2-M^2](q-l)^2l^2},
\end{equation}
where integrations over loop momenta $k$ and $l$
are performed in $D$-dimensions. Here and
below in the paper we omit the ``casual'' $i0$ in propagators for brevity.
We will expand this integral in large $M$, i.e. in small $q^2/M^2$.

Let us first obtain the part of the complete expansion proportional
to $1/(q^2)^{2\epsilon}$.  Then we will construct also 
the direct expansions for the remaining parts proportional to the factors
$1/(M^2q^2)^{\epsilon}$
and $1/(M^2)^{2\epsilon}$ allowed by power counting of non-integer
dimensions due to non-zero $\epsilon$. 
One should construct direct expansions
for all parts with different non-integer
dimensions of external parameters. This is the essence of the method. 

To get the part proportional to $1/(q^2)^{2\epsilon}$
we expand the integrand directly in large $M$.
Or, in other
words, we expand the massive propagator in the Taylor series in 
momenta $k$ and $l$
\begin{equation}
\label{largeM1a}
     T_{k,l}\left[\frac{1}{(k-l)^2-M^2}\right]=\frac{-1}{M^2}
\sum_{j=0}^{\infty} \frac{(k-l)^{2j}}{M^{2j}}.
\end{equation}
Then we get
\begin{equation}
\label{largeM2}
I_2^{(1)}= \int d^Dk~d^Dl\frac{1}
{(k-q)^2k^2(q-l)^2l^2}T_{k,l}\left[\frac{1}{(k-l)^2-M^2}\right].
\end{equation}
This direct expansion generates ultraviolet divergences, but dimensional
regularization makes them integrable. 
Hence this is the correct expansion, i.e.
it reproduces the corresponding expansion of the exact result 
in large $M$ (one first factorizes $1/M^2$ in the exact
result and then expands the rest in small $q^2/M^2$). 
By power counting of the non-integer dimensions due to 
non-zero $\epsilon$ one finds that the integrals in 
(\ref{largeM2}) are indeed proportional to $1/(q^2)^{2\epsilon}$
since they depend only on one external parameter $q$.

The complete expansion for the original integral
$I_2$ contains also terms invisible for the direct 
expansion (\ref{largeM2}).
In particular the part
proportional to $1/(M^2)^{\epsilon}$ 
is invisible in this
expansion because of the basic property of dimensional regularization
(\ref{drprop}) which in this concrete case reads
$1/(M^2)^{\epsilon +n} =0$ for $M=\infty$.
To get this part
one can make the substitution
$k^{\mu}\equiv MK^{\mu}$ in the original integral $I_2$ where $K^{\mu}$
is some new momentum.  
Then the invisible for the large $M$ expansion factor
$1/(M^2)^{\epsilon}$
is explicitly factorized by the integration measure 
$d^Dk=M^Dd^DK$ and the integral takes the form
\begin{equation}
\label{largeM1b}
I_2 =\frac{M^D}{(M^2)^3} \int\frac{d^DK~d^Dl}
{(K-q/M)^2K^2[(K-l/M)^2-1](q-l)^2l^2}.
\end{equation}
In this expression one can directly  expand the integrand in large $M$
getting the necessary contribution proportional to $1/(M^2)^{\epsilon}$.
(Then one can return to old variables
making the substitution $ K^{\mu}=k^{\mu}/M$.)
One can see that the same direct expansion can be obtained from
the original integral (\ref{largeM1}) by expanding the propagators 
carrying the momentum $k$ in the Taylor
series in momenta $q$ and $l$. Thus one gets the second contribution 
\begin{equation}
\label{largeM2a}
I_2^{(2)}= \int d^Dk~d^Dl \frac{1}{k^2(q-l)^2l^2}
T_q\left[\frac{1}{(k-q)^2}\right]
 T_l\left[\frac{1}{[(k-l)^2-M^2]}\right],
\end{equation}
where $T_q$ and $T_l$ are the operators generating the Taylor series
in $q$ and $l$ correspondingly.
Here the integrals over $k$ and $l$ factorize.
The integrals over $k$
depend only on one external parameter $M$ and are proportional
to $1/(M^2)^{\epsilon}$. The integrals over $l$ are proportional
to $1/(q^2)^{\epsilon}$. 
Thus we get the desired contribution
proportional to $1/(M^2q^2)^{\epsilon}$. 
One should also take into account
that there is a symmetrical contribution of this type. It can be
obtained from the original integral (\ref{largeM1}) if one expands
the propagators carrying the momentum $l$ 
in the Taylor series in $q$ and $k$. So the contribution $I_2^{(2)}$
comes with the factor 2 to the complete expansion.

One more way to get the same contribution $I_2^{(2)}$ is via 
the substitutions $q^{\mu}=\lambda Q^{\mu}$ and $k^{\mu}=\lambda
K^{\mu}$, where $\lambda$ is an arbitrary parameter.
Hence the factor $1/\lambda^{2\epsilon}$ is explicitely factorized by
the integration measure $d^Dk=\lambda^D d^DK$.
Then one can directly expand the integrand in $\lambda$ 
and see that remaining integrals are proportional to the necessary factor
$1/(\lambda Q)^{2\epsilon}$. 
Substitutions of this type can be useful in expansions of more
complicated integrals.

We have obtained two contributions (\ref{largeM2}) and (\ref{largeM2a}) to
the complete expansion of $I_2$. But the part proportional to
$1/(M^2)^{2\epsilon}$ is still invisible in these equations.
To get this part let us expand the integrand of the original integral
(\ref{largeM1})
directly in the Taylor series in the external momentum q. Then we
get the third contribution 
\begin{equation}
\label{largeM3}
I_2^{(3)}=\int\frac{d^Dk~d^Dl}
{k^2[(k-l)^2-M^2]l^2}T_q\left[\frac{1}{(k-q)^2(q-l)^2}\right].
\end{equation}
It is clear from power counting of non-integer dimensions
due to non-zero $\epsilon$ that integrals 
in the above equation 
are proportional to the desired factor $1/(M^2)^{2\epsilon}$ since
they depend only  on one external parameter $M$.

Summing three contributions (\ref{largeM2}),
(\ref{largeM2a}) and (\ref{largeM3}) we get the resulting expansion
\begin{equation}
\label{largeM4}
I_2=\int d^Dk~d^Dl\frac{1}
{(k-q)^2k^2(q-l)^2l^2}T_{k,l}\left[\frac{1}{(k-l)^2-M^2}\right]+
\end{equation}
\[
2\int d^Dk~d^Dl \frac{1}{k^2(q-l)^2l^2} T_q\left[\frac{1}{(k-q)^2}\right]
 T_l\left[\frac{1}{[(k-l)^2-M^2]}\right]+
\]
\[
\int\frac{d^Dk~d^Dl}
{k^2[(k-l)^2-M^2]l^2}T_q\left[\frac{1}{(k-q)^2(q-l)^2}\right].
\]
One can ask if there are  other non-perturbative
terms in the complete expansion for the original integral $I_2$
which are invisible in the above direct expansions, e.g. terms
proportional to $(q^2)^{\epsilon}/(M^2)^{3\epsilon}$. Such 
non-minimal terms (i.e. terms containing factors 
$1/(M^2)^{l\epsilon}$ or $1/(q^2)^{l\epsilon}$
 where $l$ is larger then the number of loops
of the integral)  could be present by power counting. 
But if one tries to construct
a direct expansion of  $I_2$ which would generate
such terms then the result is zero, which means that such terms
are absent.
In the
case of the threshold expansion the terms of this non-minimal type appear
for two loop propagator diagrams.

The result (\ref{largeM4}) agrees with the prescription for the 
large-mass expansion
of Feynman integrals explicitely formulated in
\cite{bdst},\cite{lrv}. The essential point of the present paper 
is the compact derivation of this expansion.

We will apply now the method to the case of the
threshold expansion of Feynman integrals.
Let us consider a scalar one loop vertex 
integral containing two massive propagators with a mass $m$ and one massless
propagator. This integral appears e.g. in the form factor for the 
decay of the virtual
photon into a quark-antiquark pair
$\gamma^{*}(q)\rightarrow \bar{Q}(p_1)Q(p_2)$ with momenta
$p_1^2=p_2^2=m^2$, $q=p_1+p_2$. One defines also the relative momentum 
$p=(p_1-p_2)/2$ with $p \cdot q=0$.
The threshold region is characterized by
\begin{equation}
     y\equiv m^2 -\frac{q^2}{4}=p^2,~~~ |y|~\ll~q^2
\end{equation}
and one considers the expansion in small $y$, i.e. in small
$y/q^2$. The corresponding integral is
\begin{equation}
\label{threshold1}
I_3 \equiv \int\frac{d^Dk}{[(k+q/2)^2-m^2][(k-q/2)^2-m^2](k-p)^2}=
\end{equation}
\[
 \int\frac{d^Dk}{(k^2+qk-y)(k^2-qk-y)(k-p)^2}.
\]
The expansion for this integral was obtained in \cite{bs}
but its consistent derivation was missing, see also \cite{s}.

Below we give the derivation of this expansion using the method.
One should again construct direct expansions for all possible
terms with different non-integer dimensions of external
parameters.
Let us first get the part proportional to
$1/(q^2)^{\epsilon}$. 
We directly expand the integrand in the Taylor series in $y$ and $p$.
All terms are integrable due to dimensional regularization
and one correctly
reproduces the expansion of the exact result for the integral 
$I_3$ in small $y$. 
The leading term of this expansion is
\begin{equation}
\label{threshold2}
I_3^{(1)}=\int\frac{d^Dk}{(k^2+qk)(k^2-qk)k^2}=
-i\pi^{D/2}\left(\frac{4}{q^2}\right)^{1+\epsilon}\frac{1}{2\epsilon}
\frac{\Gamma(1+\epsilon)}{1+2\epsilon}.
\end{equation}
One can also calculate the higher order terms in $y$ 
using e.g. Feynman parameters. 

But one can see by power counting that the
complete result can contain also the part proportional to $1/y^\epsilon$.
It is invisible for the small $y$ expansion due to the basic
property of dimensional regularization (\ref{drprop}) which now
reads $1/y^{\epsilon+n}=0$ for $y=0$.
To get this part we should construct the direct expansion in large
$q$. 
For this purpose it is convenient
to go to the frame $q=(q_0, \vec{0}), p=(0,\vec{p})$. Then the integral
 is
\begin{equation}
\label{threshold3}
I_3= \int\frac{d^Dk}{(k^2+q_0k_0-y)(k^2-q_0k_0-y)(k-p)^2}.
\end{equation}
Let us make the substitution $q_0k_0\equiv r_0$.
The integral takes the form
\begin{equation}
\label{threshold4}
I_3=\frac{1}{q_0}\int\frac{dr_0 d^{D-1}\vec{k}}
{(\frac{r_0^2}{q_0^2} -\vec{k}^2+r_0-y)(\frac{r_0^2}{q_0^2}-\vec{k}^2-r_0-y)
(\frac{r_0^2}{q_0^2}-(\vec{k}-\vec{p})^2)}.
\end{equation}
We can now expand the integrand directly in large $q_0$, i.e. in small
$r_0^2/q_0^2$. 
The resulting
integrals depend only on one external parameter $y$ producing thus
the necessary part proportional to $1/y^{\epsilon}$. 
The leading term of this expansion reads
\begin{equation}
\label{threshold5}
I_3^{(2)}=\frac{1}{q_0}\int\frac{dr_0 d^{D-1}\vec{k}}
{(-\vec{k}^2+r_0-y)(-\vec{k}^2-r_0-y)(-(\vec{k}-\vec{p})^2)}=
\end{equation}
\[
i\pi^{D/2}\frac{1}{y^{\epsilon+1/2}\sqrt{q^2}}
\frac{\sqrt{\pi}\Gamma(\epsilon+1/2)}{2\epsilon}.
\]
Note that higher order terms in the expansion in large $q_0$ give zero
since positive powers of $r_0^2$  produce (after integrations over $r_0$) 
massless tadpoles  which are zero 
in dimensional regularization.

The exact result for the integral $I_3$
can be obtained by using e.g. Feynman parameters. Then one can
see that the sum of the direct expansions (\ref{threshold2}) and 
(\ref{threshold5}) indeed reproduces
the expansion of the exact result 
\begin{equation}
\label{threshold6}
 \frac{1}{i\pi^{D/2}}I_3=
\frac{1}{y^{1+\epsilon}}
\frac{\Gamma(\epsilon)}{2}~_2 F_1\left(\frac{1}{2},1+\epsilon,\frac{3}{2};
-\frac{q^2}{4y}\right)=
\end{equation}
\[
 -\left(\frac{4}{q^2}\right)^{1+\epsilon}
 \frac{1}{2\epsilon}\sum_{j=0}^{\infty}
 \frac{\Gamma(1+\epsilon+j)}{1+2\epsilon+2j}~\frac{1}{j!}
 \left(-\frac{4y}{q^2}\right)^j
+\frac{1}{y^{\epsilon+1/2}}\frac{1}{\sqrt{q^2}}
\frac{\sqrt{\pi}\Gamma(\epsilon+1/2)}{2\epsilon}.
\]
In this way one obtains
the threshold expansion by  the method of direct expansions.

The author is grateful to the collaborators of the Institute for
theoretical particle physics of the University of Karlsruhe, where
the part of this work was done, for hospitality.
The author gratefully acknowledges the support 
by Volkswagen Foundation
under contract No. I/73611.

\end{document}